\documentclass[twocolumn]{aastex631}

\usepackage[]{mathtools}
\usepackage{bm}
\usepackage{graphicx}
\usepackage{subfigure} 
\usepackage{float}
\usepackage{verbatim}
\usepackage{booktabs}
\usepackage{multirow}
\usepackage{color}
\usepackage{mathtools}

\begin{document}

\title{Accumulating errors in tests of general relativity with gravitational waves: \\ overlapping signals and inaccurate waveforms}

\author[0000-0002-3033-6491]{Qian Hu}
%\author{Qian Hu}
\email{q.hu.2@research.gla.ac.uk}
\author[0000-0002-6508-0713]{John Veitch}
%\author{John Veitch}
\email{John.Veitch@glasgow.ac.uk}
\affiliation{Institute for Gravitational Research, School of Physics and Astronomy, University of Glasgow, Glasgow, G12 8QQ, United Kingdom}

\begin{abstract}
Observations of gravitational waves (GWs) from compact binary coalescences provide powerful tests of general relativity (GR), but systematic errors in data analysis could lead to incorrect scientific conclusions. This issue is especially serious in the third-generation GW detectors in which the signal-to-noise ratio (SNR) is high and the number of detections is large.
In this work, we investigate the impacts of overlapping signals and inaccurate waveform models on tests of GR. We simulate mock catalogs for Einstein Telescope {and Cosmic Explorer} and perform parametric tests of GR using waveform models with different levels of inaccuracy.
We find the systematic error {in non-GR parameter estimates} could accumulate toward a false deviation from GR when combining results from multiple events, {although a bayesian model selection analysis may not favour a deviation.}
%even though data from most events prefers GR.
Waveform inaccuracies contribute most to the systematic errors, but {multiple overlapping signals} could magnify the effects of systematics due to the incorrect removal of signals.
We also point out that testing GR using selected ``golden binaries'' with high SNR is even more vulnerable to false deviations from GR. The problem of error accumulation is universal; we emphasize that it must be addressed to fully exploit the data from third-generation GW detectors, and that further investigations, particularly in waveform accuracy, will be essential.
\end{abstract}
%\maketitle
\section{\label{sec1} Introduction}
The observation of gravitational waves (GWs) from compact binary coalescences (CBCs) provides an ideal means of testing of general relativity (GR) in the strong-field regime~\citep{LIGOScientific:2016lio,LIGOScientific:2017ycc,LIGOScientific:2017bnn,LIGOScientific:2018dkp,LIGOScientific:2019fpa,LIGOScientific:2020tif,LIGOScientific:2021sio}. The latest GW event catalogs contain nearly 100 CBC events~\citep{LIGOScientific:2021djp, LIGOScientific:2021usb}), based on which various tests of GR have been performed~\citep{LIGOScientific:2020tif,LIGOScientific:2021sio}. No concrete evidence of a deviation from GR has been found yet, {but unprecedented constraints have been placed on possible violations of the theory.}  
In the coming decades, the third-generation (3G) ground-based GW detectors (i.e. the Einstein Telescope~\citep{Punturo:2010zz}) and Cosmic Explorer~\citep{reitze2019_CosmicExplorerContribution}) are expected to detect $\mathcal{O}(10^5)$ CBC events per year, with signal-to-noise ratio (SNR) up to thousands~\citep{maggiore2020_ScienceCaseEinstein,himemoto2021_ImpactsOverlappingGravitationalwave,samajdar2021_BiasesParameterEstimation,relton2021_ParameterEstimationBias,oguri2018_EffectGravitationalLensing}. Since the statistical uncertainty of parameter estimates shrinks when the SNR increases, and when a catalog of events are combined, observations from 3G GW detectors are expected to be able to obtain much tighter constraints on gravity theories.

However, this inspiring prospect of an enlarged detection catalog and higher SNRs brings with it many difficulties in data analysis. For the purpose of testing GR (and any other theories), one needs to ensure that the systematic errors are small, so that the analysis will not favor the wrong theory and cause a false alarm (or false dismissal). Parameterized tests of GR~\citep{meidam2018_ParameterizedTestsStrongfield} suffer from the same problems as parameter estimation (PE) in general, which has been investigated in many works, (e.g. \citep{cutler2007_LISADetectionsMassive,antonelli2021_NoisyNeighboursInference}. 
For instance, inaccurate waveform models may have already caused some tensions in current GW observations~\citep{hu2022_AssessingModelWaveform, Williamson:2017evr} and are expected to be more important in future high SNR detections~\citep{cutler2007_LISADetectionsMassive,PhysRevD.103.124015,purrer2020_GravitationalWaveformAccuracy}. Additionally, the 3G detectors {with their improved low-frequency sensitivity} are able to observe multiple signals at the same time. Detected overlapping signals can not be perfectly removed from the data, and could have non-negligible impact on PE when the merger times of overlapping signals are close~\citep{himemoto2021_ImpactsOverlappingGravitationalwave,samajdar2021_BiasesParameterEstimation,relton2021_ParameterEstimationBias,antonelli2021_NoisyNeighboursInference,Pizzati:2021apa, Janquart:2022nyz}. The undetected overlapping signals, i.e., the signals that are too faint to be detected may also contribute to the systematic error~\citep{antonelli2021_NoisyNeighboursInference,Reali:2022aps}. These errors are inevitable in 3G detectors, and repeated biased estimations for each event might end up with a wrong conclusion in the catalog-level analysis \citep{PhysRevD.105.L061301,moore2021_TestingGeneralRelativity}. 

{Aforementioned works mainly focus on case studies for single events, or include only one type of systematic error. In this work, we aim to perform a more comprehensive investigation on systematic errors at the catalog level, including interactions between different types of systematics.}  We perform parameterized post-Newtonian (PPN) coefficient tests~\citep{Mishra:2010tp,Cornish:2011ys,Li:2011cg} with our simulated event catalogs and inaccurate waveforms. Our simulations show that systematic errors can accumulate, and could lead to an incorrect measurement of deviation from GR when results from multiple events are combined. We find overlapping signals could magnify the effects of waveform systematics because of their imperfect subtraction from the data . Even worse, we find that the selected high-SNR events without known overlapping signals (so-called ``golden events'', which have been examined for GR tests in e.g. \citep{PhysRevD.94.021101}) may be more vulnerable to biased conclusions.

{This paper is organized as follows. We introduce our methodology in Sec.~\ref{sec2}, including the Fisher matrix formalism for error prediction in Sec.~\ref{sec2a}, configurations of parameter estimation, waveforms and PPN tests in Sec.~\ref{sec2b}, catalog simulation and overlapping signals in Sec.~\ref{sec2c}, and methods of combining results in Sec.~\ref{sec2d}. Results are given in Sec.~\ref{sec3}. We first demonstrate selected example events in Sec.~\ref{sec3a}, and then move on to the catalog level tests in in Sec.~\ref{sec3b} and Sec.~\ref{sec3c}. Conclusions and discussions are in Sec.~\ref{sec4}.
}

\section{Systematic biases in PPN tests \label{sec2}}

\subsection{Estimating systematic errors\label{sec2a}}
The generic formalism we use for estimating systematic errors in PE was first proposed in \citet{cutler2007_LISADetectionsMassive} and then generalized and {validated against full PE} by \citet{antonelli2021_NoisyNeighboursInference}. Let $\bm{\theta}$ be the parameters of a GW signal $h(\bm{\theta})$ (which may include more than one source, in the case of overlapping signals). The frequency domain data from a GW detector, denoted $d(\bm{\theta})$ is given by
\begin{equation}
    d(\bm{\theta}) = h(\bm{\theta}) + n,
\end{equation}
where $n$ is noise. Under the assumption that this noise is stationary and Gaussian, the likelihood for GW PE is
\begin{equation}
    L(\bm{\theta}) \propto e^{-\frac{1}{2} (d-h|d-h)} = e^{-\frac{1}{2} (n|n)},
\end{equation}
where $(\dots | \dots)$ is the inner product~\citep{finn1992_DetectionMeasurementGravitational}, defined as
\begin{equation}
    (a | b) = 4\Re \int_{0}^{\infty} \frac{a^*(f) b(f)}{S_{\mathrm{n}}(f)}df,
\end{equation}
where $*$ means complex conjugate, and $\Re$ denotes the real part.  $S_{\mathrm{n}}(f)$ is the {noise} power spectral density (PSD) of the detector. The optimal SNR is $\rho = \sqrt{(h|h)}$. For more than one data streams, the inner product's definition should be replaced by the sum of inner products calculated individually by each data stream.

Consider a maximum likelihood estimator (which is equivalent to Bayesian estimation with flat priors), the maximum point $\bm{\theta}_\mathrm{ML}$ satisfies
\begin{equation}
    \label{eq:ml}
    \partial_i \ln L \mid_{\bm{\theta}= \bm{\theta}_\mathrm{ML}} = (\partial_i h | d-h) \mid_{\bm{\theta}= \bm{\theta}_\mathrm{ML}}=0,
\end{equation}
where $\partial_i$ denotes the derivative with respect to the i'th parameter. The data $d$ is known, but real parameter $\bm{\theta}_\mathrm{real}$ and the GW signal in the detector $h(\bm{\theta}_\mathrm{real})$ is unknown. In practice, they are replaced by a waveform model $h_\mathrm{m}(\bm{\theta}_\mathrm{ML})$. By doing this, errors are introduced to $d-h$: 
\begin{align}
    %\begin{aligned}
    %    &d-h_\mathrm{m}(\bm{\theta}_\mathrm{ML}) \nonumber \\ 
    %    &= d - h(\bm{\theta}_\mathrm{real}) + h(\bm{\theta}_\mathrm{real}) - h_\mathrm{m}(\bm{\theta}_\mathrm{real})  + h_\mathrm{m}(\bm{\theta}_\mathrm{real}) - h_\mathrm{m}(\bm{\theta}_\mathrm{ML})  \nonumber  \\
    d-h = n + \delta H + \Delta \theta^j \partial_j h_\mathrm{m}. \label{eq:dminush}
    %\end{aligned}
\end{align}
The first term $n$ is what $d-h$ is supposed to be: the noise in the detector. The second term $\delta H = h(\bm{\theta}_\mathrm{real}) - h_\mathrm{m}(\bm{\theta}_\mathrm{real})$ is the excess strain which represents the difference between real signal(s) in the data and the model used to subtract signals. Inaccurate waveforms and overlapping signals can both contribute to this term. The third term comes from {the imperfect measurement of signal parameters due to statistical noise, and is given by the} linear expansion of $h_\mathrm{m}(\bm{\theta}_\mathrm{real}) - h_\mathrm{m}(\bm{\theta}_\mathrm{ML})$, where $\Delta \theta^j$ is the {statistical} error of the j'th parameter from the maximum likelihood estimator, and we adopt Einstein notation to indicate the sum over parameters. Substituting Eq.~\ref{eq:dminush} into Eq.~\ref{eq:ml} and approximate all derivatives at $\bm{\theta}_\mathrm{ML}$, and we get
\begin{equation}
    \label{eq:errors}
    \Delta \theta^i \approx (\bm{\Gamma}^{-1})^{ij} (\partial_j h_\mathrm{m} | n + \delta H) = \Delta \theta^i_\mathrm{stat} + \Delta \theta^i_\mathrm{sys},
\end{equation}
where $\bm{\Gamma}_{ij} = (\partial_i h_\mathrm{m} | \partial_j h_\mathrm{m})$ is the Fisher matrix~\citep{cutler1994_GravitationalWavesMerging}. $\Delta \theta^i_\mathrm{stat} = (\bm{\Gamma}^{-1})^{ij} (\partial_j h | n)$ is the error induced by the detector noise.  $<\Delta \theta^i_\mathrm{stat}>=0$, so the maximum likelihood estimator is unbiased if $\delta H=0$; and $<\Delta \theta^i_\mathrm{stat}\Delta \theta^j_\mathrm{stat}> = (\bm{\Gamma}^{-1})^{ij}$, which is consistent with the Fisher matrix formalism.
%used in detectability forecast
The $\Delta \theta^i_\mathrm{sys} = (\bm{\Gamma}^{-1})^{ij} (\partial_j h_\mathrm{m} | \delta H)$ is the systematic error. Any effect that contributes to $\delta H$ could be a source of systematic bias in PE. 
We will use $\sqrt{(\bm{\Gamma}^{-1})^{ii}}$ as statistical uncertainty and $\Delta \theta^i_\mathrm{sys}$ as the predicted systematic error.

\subsection{PPN formalism, choices of parameters and waveforms \label{sec2b}}
The test of parameterized post-Newtonian coefficients is a generic formalism for finding deviations from GR, initially proposed by \citet{Mishra:2010tp} and further developed for application with Bayesian inference~\citep{Li:2011cg}, and later applied to catalogs of real GW observations, most recently in \citet{LIGOScientific:2021sio}. We use the waveform model \texttt{IMRPhenomPv2}~\citep{Husa:2015iqa,Khan:2015jqa}, whose phase is characterized by a set of parameters $\{p_i \}$, including inspiral phase parameters $\{\phi_0,\dots, \phi_7  \}$ and $\{\phi_{5l}, \phi_{6l}  \}$, phenomenological coefficients $\{\beta_0,\dots, \beta_3  \}$, and merger-ringdown parameters $\{\alpha_0,\dots, \alpha_5  \}$. Deviations $p_i \rightarrow (1+\delta \hat{p}_i)p_i$ are introduced as the violations of GR; $\delta \hat{p}_i=0$ reproduces GR. In this framework, testing GR is reduced to estimating the testing parameters $\delta \hat{p}_i$. Although a specific modified gravity theory could bring deviations in more than one testing parameter, previous works have shown that including one testing parameter at once is enough to detection violations. In fact it can be more efficient to find violations from GR this way because it avoids the correlations between testing parameters and GR parameters~\citep{meidam2018_ParameterizedTestsStrongfield,Sampson:2013lpa}. In this work, we choose $\delta \hat{\phi}_0$ as the example testing parameter. We assume GR is the correct theory and focus on whether the PPN test falsely indicates deviations of GR.

{We restrict our Fisher matrix analysis to a subset of the full signal parameters, to avoid computational issues. Parametrized deviations of the type we consider have a direct effect on the phasing of the signal, so in addition to $\delta \hat{\phi}$ we must include the other parameters that do the same: chirp mass $\mathcal{M}$ and mass ratio $q$, as well as the time of coalescence $t_c$. The full 6-dimensional space of spin configurations is known to bring ill-conditioned Fisher matrices~\citep{Borhanian:2022czq} due to correlations between parameters, and because of the prior bounds on angular parameters results can be misleading even when they can be computed. We therefore use only the effective spin $\chi_\mathrm{eff}$ to capture the dominant effect of (aligned) spin on the waveforms. We include this by forcing the two aligned spin components to contribute equally to $\chi_\mathrm{eff}$, which allows us to treat it as a single parameter.
We neglect to include extrinsic parameters in the Fisher matrix, effectively assuming they are measured precisely. Since these do not have a frequency-dependent effect on the phase, we do not expect them to be highly correlated with the intrinsic parameters.}
Our choice captures the parameters that appears in the leading PN term and the corresponding PPN modifications, as well as the decisive parameter in the analysis of overlapping signals, $t_c$.
%Therefore, as a qualitative investigation, we only consider {four} GR parameters (chirp mass $\mathcal{M}$, mass ratio $q$, {effective spin $\chi_\mathrm{eff}$ and coalescence time $t_\mathrm{c}$) in Fisher matrix. We assume aligned spins and enforce two components have same contribution to the effective spin. $\chi_\mathrm{eff}$ can then be treated as one parameter, not a combination of many.}
{ Other parameters are randomly generated (details in Sec.~\ref{sec2c}) but are treated as perfectly known. Setting parameters to their injection values excludes their contributions to both statistical and systematic errors in PE. For instance, if we removed the effective spin from our calculation, we would obtain tighter statistical and systematic errors because its correlation with mass parameters and the testing parameter is removed~\citep{Berti:2004bd}. 
Considering realistic PE in the future in which all parameters are included, correlation between parameters may make posteriors wider and systematic bias larger. However, due to the linear expression in Eq.~\ref{eq:errors}, we expect the two changes are proportional and our conclusion will not change significantly under this simplification.}

{We induce a non-zero $\delta \hat{\beta}_2$ to mimic inaccurate waveform models based on the following considerations. To reduce potential correlations with $\delta \hat{\phi}_0$, we exclude testing parameters for the inspiral stage. Correlation between the testing parameter and the waveform systematic parameter may undermine the generality of the illustration. To make sure the testing parameter has enough influence on the waveform, we do not choose parameters for the merger-ringdown stage which only includes the last few cycles. Therefore, we look for parameters in the intermediate region which is described by $\delta \hat{\beta}_i$~\citep{Khan:2015jqa}. $\delta \hat{\beta}_0$ and $\delta \hat{\beta}_1$ bring global phase shift and time shift in this region respectively, so $\delta \hat{\beta}_2$ is the dominant testing parameter that encodes physical (frequency-dependent) modifications. }

\begin{figure}
    \includegraphics[width=0.47\textwidth]{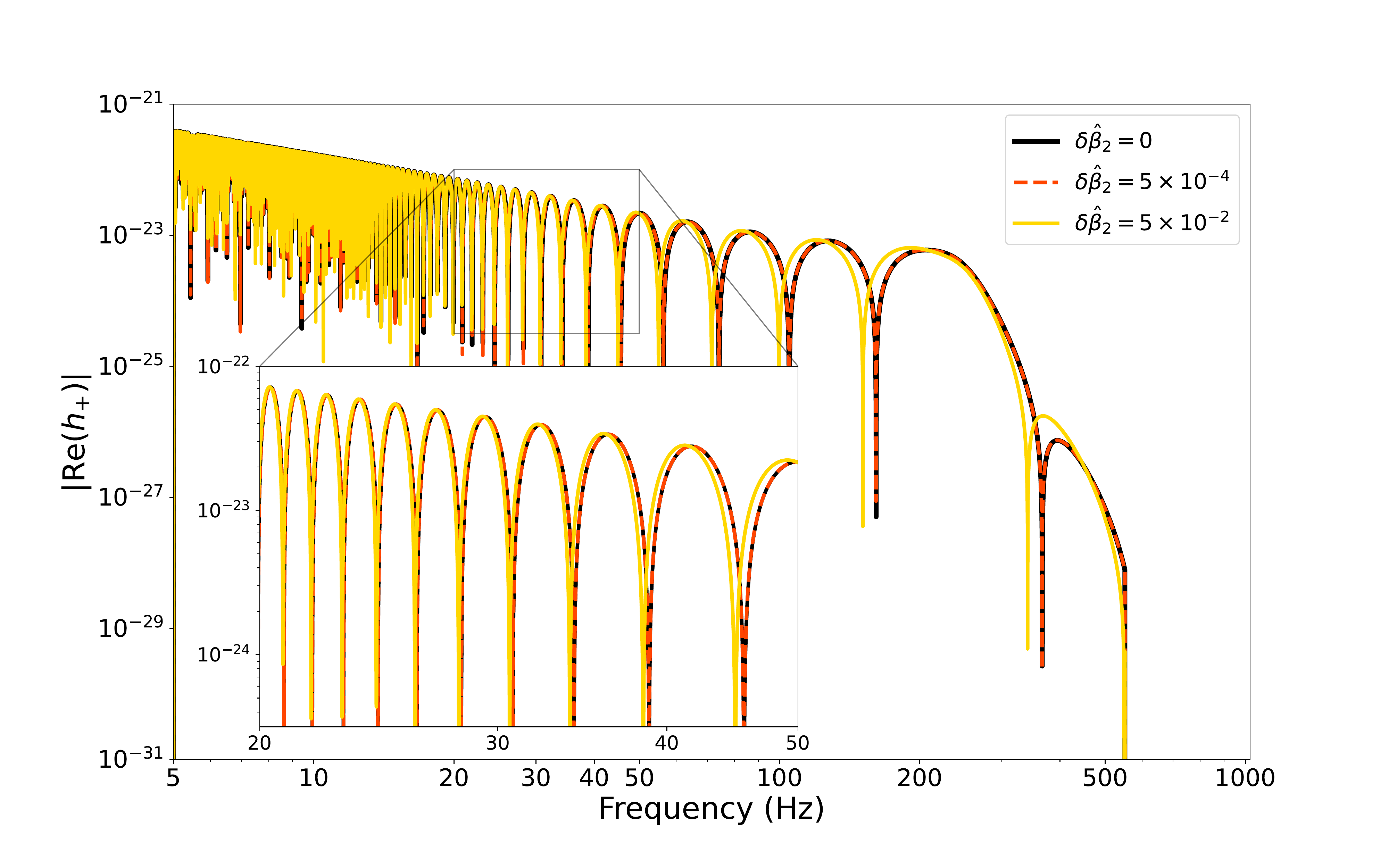}
    \centering
    \caption{\label{pic:wf_compare} {The absolute value of real part of plus polarization from a non-spinning BBH with $\mathcal{M}_c = 30.69~\mathrm{M_\odot}$, $q=0.88$, in frequency domain. Waveforms with different $\delta \hat{\beta}_2$ are shown in different color and linestyles. The intermediate region of this system starts around 50\,Hz, which is consistent with where waveform difference appears in the plot. }} 
\end{figure}

We assume $\delta \hat{\beta}_2=0$ is our model waveform, while the ``real'' waveform could have $\delta \hat{\beta}_2=0$, $5\times 10^{-2}$, or $5\times 10^{-4}$. The first case means our model waveform is perfect, and all systematic errors will come from overlapping signals. The second case generates waveform mismatches around $10^{-4} - 10^{-3}$, which corresponds to the current waveform accuracy~\citep{pratten2021_ComputationallyEfficientModels,ossokine2020_MultipolarEffectiveonebodyWaveforms}. The last case produces mismatches around $10^{-7} - 10^{-6}$ and corresponds to the expectations for future waveform accuracy~\citep{purrer2020_GravitationalWaveformAccuracy,hu2022_AssessingModelWaveform}. {A comparison of the three types of waveforms is shown in Fig.~\ref{pic:wf_compare}. We show an example of a non-spinning BBH merger with $\mathcal{M}_c = 30.69~\mathrm{M_\odot}$ (in the detector frame), and $q=0.88$  whose intermediate region starts around 50Hz. The mismatches are $3\times 10^{-7}$ and $2\times 10^{-3}$ between $\delta \hat{\beta}_2=0$ and $\delta \hat{\beta}_2=5\times 10^{-4}, 5\times 10^{-2}$, respectively. }

The excess strain from inaccurate waveforms can be written as 
\begin{equation}
    \delta H_{\mathrm{wf}} = h(\bm{\theta}_\mathrm{real}) \mid_{\delta \hat{\beta}_2\neq 0} - h(\bm{\theta}_\mathrm{real})\mid_{\delta \hat{\beta}_2=0}     %\delta H_{\mathrm{wf}} = h(\bm{\theta}_\mathrm{real}) \mid_{\delta \hat{\beta}_2=\{0, 5\times 10^{-2}, 5\times 10^{-4} \} } - h(\bm{\theta}_\mathrm{real})\mid_{\delta \hat{\beta}_2=0}
\end{equation}
We can use the approximation $h(\bm{\theta}_\mathrm{real}) - h_\mathrm{m}(\bm{\theta}_\mathrm{real}) \approx h(\bm{\theta}_\mathrm{ML}) - h_\mathrm{m}(\bm{\theta}_\mathrm{ML}) $, as the error would be a higher order term. 

\subsection{Overlapping signals and mock catalogs\label{sec2c}}
When multiple signals come into data, they may have impacts on the analysis of each other~\citep{himemoto2021_ImpactsOverlappingGravitationalwave,samajdar2021_BiasesParameterEstimation,relton2021_ParameterEstimationBias,antonelli2021_NoisyNeighboursInference}. It is known that the correlation between signals is not strong unless the merger times are very close {(typically $<1$s)}; in this work we regard two signals as ``overlapping'' only if the merger time difference $|\Delta t|<4$s{, which captures the most influential neighbours of a signal}. 

Overlapping signals can be classified into two types: detected signals and undetected signals (confusion signals). The former is strong enough to be detected and should be subtracted from data in the analysis for other signals (or, the ``main'' signal)\footnote{{It is also possible to do a joint parameter estimation for all existing signals, see \citet{Janquart:2022nyz}.}}. The latter, however, is too faint to be recognized by the {detection pipeline} and may have an unnoticed impact on PE. In this work, the {network} SNR threshold for detection is set to 8, under which GWs are assumed to be undetected. 

If a signal is detected, it will still contribute excess strain since we cannot perfectly remove it from the data. The excess strain after imperfect removal is
\begin{equation}
    \label{eq:do}
    \delta H_{\mathrm{DO}} = h'(\bm{\theta}_\mathrm{real}) - h_\mathrm{m}'(\bm{\theta}_\mathrm{ML}) \approx \Delta \theta'^{i} \partial_i h'_m + \delta H'_{\mathrm{wf}},
\end{equation}
where $'$ denotes variables of the detected overlapping signal. The first term arises from the inaccurate estimation of parameters for the overlapping signal, which is random since the error is partly caused by the random noise, {although} {other factors, such as waveform inaccuracies and overlapping signals also contribute to it. As a conservative estimation  and following \citet{antonelli2021_NoisyNeighboursInference}, we ignore waveform systematic errors in $\Delta \theta'^{i}$ (i.e., assuming $\Delta \theta'^{i}$ is merely caused by noise, which tends to underestimate it), and adopt the lowest order approximation for its correlation with the main signal.} Substituting it into Eq.~\ref{eq:errors}, one obtains the covariance of the first term in {the systematic error Eq.~\ref{eq:do}}
\begin{equation}
    \label{eq:do1}
    <\Delta \theta^i_{\mathrm{DO1}} \Delta \theta^j_{\mathrm{DO1}}>  = \left( \bm{\Gamma}^{-1} \bm{\Gamma}^{-1}_{\mathrm{mix}} \bm{\Gamma}^{'-1} (\bm{\Gamma}^{-1}_{\mathrm{mix}})^\mathrm{T}  (\bm{\Gamma}^{-1})^\mathrm{T}  \right)_{ij},
\end{equation}
where $(\bm{\Gamma}_{\mathrm{mix}})_{ij} = (\partial_i h | \partial_j h')$ encodes the correlation between two signals and $\bm{\Gamma}^{'} = (\partial_i h' | \partial_j h')$ is the Fisher matrix of the overlapping signal. The second term in Eq.~\ref{eq:do} represents the inaccurate waveform model we use to subtract signals, and can be calculated the same way as the waveform systematic, yielding $\Delta \theta^i_{\mathrm{DO2}} =  (\bm{\Gamma}^{-1})^{ij} (\partial_j h_\mathrm{m} | \delta H'_{\mathrm{wf}})$. In this work, the systematic error from detected overlapping signals is calculated as $\Delta \theta^i_{\mathrm{DO2}}$ plus a random sample drawn from a multivariate Gaussian distribution with covariance matrix Eq.~\ref{eq:do1} and zero mean. For more than one detected overlapping signal, Eq.~\ref{eq:do} can be extended by defining $h'$ as the summation of all GWs in the data~\citep{antonelli2021_NoisyNeighboursInference}, which enlarges the dimension of $\bm{\Gamma}_{\mathrm{mix}}$ and $\bm{\Gamma}^{'}$. 

The undetected overlapping signal simply contributes to systematic error by $\delta H_\mathrm{UO} = \sum_{\mathrm{undetected}} h''(\bm{\theta}_\mathrm{real})$. It is accessible in our simulation but unknown in real data analysis.

We consider BBH and BNS sources, and assume their distribution in redshift $z$ follows the analytical approximation~\citep{oguri2018_EffectGravitationalLensing}
\begin{equation}
    R_{\mathrm{GW}}(z)=\frac{a_1 e^{a_2 z}}{e^{a_3 z}+a_4} \mathrm{Gpc^{-3}yr^{-1}},
\end{equation}
which is then converted to observable event rate by multiplying a factor $\frac{1}{1+z}\frac{dV_\mathrm{c}}{dz}$. Here $V_\mathrm{c}$ is the comoving volume and we employ Planck15 cosmology~\citep{Planck:2015fie}. Note that ``observable'' GWs need to achieve an {network} SNR of 8 to be ``detectable''. $a_{\{1,2,3,4\}}$ are model parameters. We set $a_2 = 1.6, a_3=2.1, a_4 = 30$ to mimic a peak at $z\sim 2$. $a_1$ is scaled based on local merger rate given by \citet{LIGOScientific:2020kqk} ($\mathcal{R}_{\mathrm{BNS}}=320_{-240}^{+490}$ and $\mathcal{R}_{\mathrm{BBH}}=23.9_{-8.6}^{+14.3} \mathrm{Gpc}^{-3} \mathrm{yr}^{-1}$) such that $R_{\mathrm{GW}}(z=0) = \mathcal{R}_{\mathrm{BNS/BBH}}$. We choose three values for $a_1$ which corresponds to lower, median, and higher estimation of local merger rate, respectively. 

The masses of BBHs are generated by the PowerLaw $+$ Peak model in \citet{LIGOScientific:2020kqk}, while all BNS systems are set to be same: $1.45+1.4 M_\odot$, $\Lambda_1 = \Lambda_2 = 425$. {The effective spin follows the Gaussian distribution in \citet{LIGOScientific:2020kqk}, with mean of 0.06 and standard {deviation} of 0.12.} \texttt{IMRPhenomPv2\_NRTidal} \citep{Dietrich:2018uni} is used to generate BNS waveforms with the same $\delta \hat{\beta}_2$ as BBH. We will perform tests of GR with all BBH events and use BNS events as a background: BNS events are only involved in the calculation as overlapping signals. 
We assume isotropically distributed inclination and source sky direction; and uniformly distributed coalescence time, phase, and polarization angle. 

A summary of low, median, and high merger rates catalogs is shown in Tab.~\ref{tab1}. It shows that most BBH events will not have an overlapping signal near their merger time, which implies overlapping signals contribute to systematic errors less frequently than waveform systematics. {With our ET+CE configuration, the numbers of the two kinds of overlaps are close. However, if the number of detectors is less than assumed, or detector sensitivities are lower than designed, some of detected overlaps would become undetected, and vice versa.} The unnoticeable confusion background has drawn attention in recent works~\citep{Wu:2022pyg,Reali:2022aps} and needs further investigation. {Compact binaries formed by Pop III stars (which we have ignored) could also contribute to the confusion background. However, according to the model in \citet{oguri2018_EffectGravitationalLensing}, the numbers of observable Pop III binaries of B17 and K16 models per year are roughly 40000 and 180000 respectively, which is much lower than the BNS background. }

\begin{table*}[]
    \begin{ruledtabular}
\begin{tabular}{|c|cc|cc|cc|}
\hline
                        & \multicolumn{2}{c|}{\# of observable binaries}                          & \multicolumn{2}{c|}{Detected overlaps on BBH events}          & \multicolumn{2}{c|}{Undetected overlaps on BBH events}        \\ \hline
                        & \multicolumn{1}{c|}{BBH}                     & BNS                      & \multicolumn{1}{c|}{\# of overlaps} & \# (fraction) of events & \multicolumn{1}{c|}{\# of overlaps} & \# (fraction) of events \\ \hline
\multirow{4}{*}{Low}    & \multicolumn{1}{c|}{\multirow{4}{*}{56526}}  & \multirow{4}{*}{286088}  & \multicolumn{1}{c|}{0}              & 53118 (95\%)            & \multicolumn{1}{c|}{0}              & 54067 (96\%)            \\ \cline{4-7} 
                        & \multicolumn{1}{c|}{}                        &                          & \multicolumn{1}{c|}{1}              & 2847 (5.1\%)            & \multicolumn{1}{c|}{1}              & 1936 (3.5\%)            \\ \cline{4-7} 
                        & \multicolumn{1}{c|}{}                        &                          & \multicolumn{1}{c|}{2}              & 74 (0.13\%)             & \multicolumn{1}{c|}{2}              & 37 (0.066\%)           \\ \cline{4-7} 
                        & \multicolumn{1}{c|}{}                        &                          & \multicolumn{1}{c|}{3}              & 2 (0.0040\%)             & \multicolumn{1}{c|}{3}              & 1 (0.0018\%)            \\ \hline
\multirow{5}{*}{Median} & \multicolumn{1}{c|}{\multirow{5}{*}{88300}}  & \multirow{5}{*}{1144354} & \multicolumn{1}{c|}{0}              & 73200 (84\%)            & \multicolumn{1}{c|}{0}              & 76270 (87\%)            \\ \cline{4-7} 
                        & \multicolumn{1}{c|}{}                        &                          & \multicolumn{1}{c|}{1}              & 13125 (15\%)            & \multicolumn{1}{c|}{1}              & 10461 (12\%)            \\ \cline{4-7} 
                        & \multicolumn{1}{c|}{}                        &                          & \multicolumn{1}{c|}{2}              & 1093 (1.2\%)            & \multicolumn{1}{c|}{2}              & 721 (0.82\%)            \\ \cline{4-7} 
                        & \multicolumn{1}{c|}{}                        &                          & \multicolumn{1}{c|}{3}              & 67 (0.077\%)            & \multicolumn{1}{c|}{3}              & 35 (0.040\%)            \\ \cline{4-7} 
                        & \multicolumn{1}{c|}{}                        &                          & \multicolumn{1}{c|}{4}              & 2 (0.0023\%)            & \multicolumn{2}{c|}{}                                         \\ \hline
\multirow{7}{*}{High}   & \multicolumn{1}{c|}{\multirow{7}{*}{143349}} & \multirow{7}{*}{2896647} & \multicolumn{1}{c|}{0}              & 92692 (65\%)            & \multicolumn{1}{c|}{0}              & 100862 (71\%)           \\ \cline{4-7} 
                        & \multicolumn{1}{c|}{}                        &                          & \multicolumn{1}{c|}{1}              & 39450 (28\%)            & \multicolumn{1}{c|}{1}              & 34519 (24\%)            \\ \cline{4-7} 
                        & \multicolumn{1}{c|}{}                        &                          & \multicolumn{1}{c|}{2}              & 8559 (6.0\%)           & \multicolumn{1}{c|}{2}              & 5940 (4.2\%)           \\ \cline{4-7} 
                        & \multicolumn{1}{c|}{}                        &                          & \multicolumn{1}{c|}{3}              & 1208 (0.85\%)           & \multicolumn{1}{c|}{3}              & 673 (0.47\%)           \\ \cline{4-7} 
                        & \multicolumn{1}{c|}{}                        &                          & \multicolumn{1}{c|}{4}              & 131 (0.092\%)           & \multicolumn{1}{c|}{4}              & 58 (0.041\%)           \\ \cline{4-7} 
                        & \multicolumn{1}{c|}{}                        &                          & \multicolumn{1}{c|}{5}              & 20 (0.014\%)           & \multicolumn{1}{c|}{5}              & 7 (0.0049\%)            \\ \cline{4-7} 
                        & \multicolumn{1}{c|}{}                        &                          & \multicolumn{2}{c|}{}                                         & \multicolumn{1}{c|}{6}              & 1 (0.00070\%)            \\ \hline
\end{tabular}

    \caption{\label{tab1} A summary of three mock catalogs. From left to right, it shows catalog type, observable BBH and BNS per year (note this is not detectable), {and distributions of numbers of overlapping signals among BBH events. For example, in median merger rate catalog, there are 13125 detected BBH events (15\% of all detected BBH events) coming with 1 detected overlapping GW signal, and 10461 detected BBHs coming with 1 undetected overlapping GW signal.
    The overlapping signal can be BBH or BNS, and two signals are defined as overlapped if their merger time difference $\Delta t < 4$s. }}
    \end{ruledtabular}
\end{table*}

Several simplifications have been adopted in our mock catalog: we regard BNS as a background and use only BBH as the test source; we ignore neutron star-black hole (NSBH) mergers and other possible types of sources; we use an analytical merger rate that peaks at $z\sim 2$, ignoring compact binaries from Pop III stars. Our catalogs aim to generate an appropriate merger rate for the study of systematic error accumulation, rather than accurately modeling the astrophysical population. To achieve this, we also adjust the merger rate to different levels, expecting that the real situation will lie somewhere between our lowest and highest estimates. 

{Signals are injected into the 3rd generation GW detector Einstein Telescope with ET-D PSD \citep{Punturo:2010zz} {located at the Cascina site} of the current Virgo detector, and Cosmic Explorer located at the LIGO Hanford site with the sensitivity curve proposed by \citet{LIGOScientific:2016wof}. The frequency band used for the analysis is 5--2048\,Hz. }

\subsection{Combining results}\label{sec2d}

There are several ways of combining results from multiple events~\citep{zimmerman2019_CombiningInformationMultiple,isi2019_HierarchicalTestGeneral}. We employ two straightforward methods: multiplying likelihoods (equivalently, multiplying posteriors if priors are flat) and multiplying Bayes factors. The former assumes the modification parameter is the same for all events, while the latter allows the modification parameter to vary across events. 

{We assume a flat prior distribution, and that the posterior follows a multivariate Gaussian distribution with covariance matrix $\bm{\Gamma}^{-1}$ and mean $\bm{\mu}$ equal to injection values $\mathbf{\theta_\text{inj}}$ plus systematic errors $\mathbf{\Delta \theta_\mathrm{sys}}$. The statistical uncertainty of a parameter is $\sigma_i = \sqrt{(\bm{\Gamma}^{-1})_{ii}}$. We define the error ratio} {between systematic and statistical errors as}
\begin{equation}
    \mathcal{R}(\theta_i) = |\Delta \theta_{\mathrm{i,~sys}}/\sigma_i |.
\end{equation}
{We consider that} the PPN test coefficient is subject to false deviations from GR when $\mathcal{R}(\delta \hat{\phi}_0)>1$.

{In order to combine results from multiple events, one would multiply the} {posterior distributions of the testing parameter} {for each.} {Multiplication of Gaussian distributions results in another Gaussian distribution whose mean (systematic error) is a linear combination of the original means. From the first event in a catalog, we multiply the posterior of new events one by one and calculate the error ratio. Considering the arbitrary sequence of events, we permute the sequence 200 times and extract the ensemble average and 68\% confidence interval.}
%Note that, the multiplication of two Gaussian distributions is still Gaussian whose mean (systematic error) is a linear combination of two original means. Therefore, a plus systematic error and a minus systematic error could neutralize if we combine events by multiplying likelihood. {The error ratio may oscillate during accumulation, but it is unlikely to return to 0, which means } 

{Treating GR as a sub-model of the non-GR theory, Bayes factor can be calculated analytically with the Gaussian posterior~\citep{moore2021_TestingGeneralRelativity}. Denote systematic error of  $\delta \hat{\phi}_0$ as $\Delta\theta_{\mathrm{sys}}$, we have }
\begin{equation}
    L_\mathrm{GR}(\bm{\theta}_\mathrm{GR}) = L_\mathrm{nonGR}(\bm{\theta}_\mathrm{nonGR})\mid_{\delta \hat{\phi}_0=\Delta\theta_{\mathrm{sys}}}.
\end{equation}
The Bayes factor is then calculated as
\begin{equation}
    \begin{aligned}
        &\mathcal{B}^\mathrm{nonGR}_\mathrm{GR} \sim \frac{Z_\mathrm{nonGR}}{Z_\mathrm{GR}} = \frac{\int \bm{\mathrm{d}}\bm{\theta}_\mathrm{nonGR} L_\mathrm{nonGR}}{\int \bm{\mathrm{d}}\bm{\theta}_\mathrm{GR} L_\mathrm{GR}}\\
        &= \sqrt{2\pi} e^{\frac{1}{2}(\bm{\Gamma}_{\delta \hat{\phi}_0 \delta \hat{\phi}_0} - \mathbf{v}^\mathrm{T}(\bm{\Gamma}_\mathrm{GR})^{-1}\mathbf{v} ) \Delta\theta_{\mathrm{sys}}^2} \sqrt{\frac{\det \bm{\Gamma}_\mathrm{GR}}{\det \bm{\Gamma}_\mathrm{nonGR}}}, 
    \end{aligned}
\end{equation}
{where $\bm{\Gamma}_\mathrm{nonGR}$ is the Fisher matrix including the testing parameter, while $\bm{\Gamma}_\mathrm{GR}$ only includes GR parameters. $\mathbf{v}_i= (\partial h/\partial \bm{\theta}_i|\partial h/\partial\delta \hat{\phi}_0)$ represents the correlation between GR and non-GR parameters. $\bm{\Gamma}_{\delta \hat{\phi}_0 \delta \hat{\phi}_0} = (\partial h/\partial\delta \hat{\phi}_0|\partial h/\partial\delta \hat{\phi}_0)$. The exponential term in the Bayes factor accounts for the deviation of GR, while the determinant ratio term usually favors GR since modified theories introduce extra parameters to explain the data. We also note that the correlation term $\mathbf{v}^\mathrm{T}(\bm{\Gamma}_\mathrm{GR})^{-1}\mathbf{v}$ mitigates the deviation of GR. Ignoring this term may overestimate the Bayes factor (e.g., \citet{moore2021_TestingGeneralRelativity}).}
{When combining events, }{Bayes factors are numerically multiplied, with the same permutation mentioned before. {We consider a }false deviation from GR {to be} achieved when $\ln\mathcal{B}^\mathrm{nonGR}_\mathrm{GR} > 8$.}
{We reemphasize that Bayes factors are first computed for each event and then combined across the catalog, rather than calculated after different posteriors are multiplied.
This analysis should be interpreted as not assuming that the
testing parameter is the same for all events. In this sense it is less sensitive to violations of GR when there is a common underlying deviation parameter, so we would expect it to be less vulnerable to simulated false violations.
While error ratio accumulation is decided by errors from each event, Bayes factor accumulation is more sensitive to the fraction of correct analyses in the catalog. The two methods of combining results are independent and do not necessarily lead to the same conclusion. More details are given in Sec.~\ref{sec3b}. }

\section{Results\label{sec3}}

\subsection{Single events\label{sec3a}}
We first present an example event, {investigating the effect of a detected or undetected overlapping signal}. The main signal is from a BBH with $\mathcal{M}_\mathrm{c} = 32 \mathrm{M}_\odot$ (in the detector frame), $q=0.9$, $\chi_\mathrm{eff}=0.2$ and network SNR of 27. The overlapping signal is an equal mass BBH with $\mathcal{M}_\mathrm{c} = 20 \mathrm{M}_\odot$ and $\chi_\mathrm{eff}=0.1$. We scale its SNR from $\sim 26$ down to $\lesssim  8$ to make it detectable or undetectable. We vary the merger time difference (by 0.01\,s per step) and calculate the total systematic error with different waveform models. {Note that, throughout this section, the ``systematic error'' refers to that of the testing parameter $\delta \hat{\phi}_0$, and is denoted as $\Delta \theta_{\mathrm{sys}}$.  }

\begin{figure*}
    \includegraphics[width=0.97\textwidth]{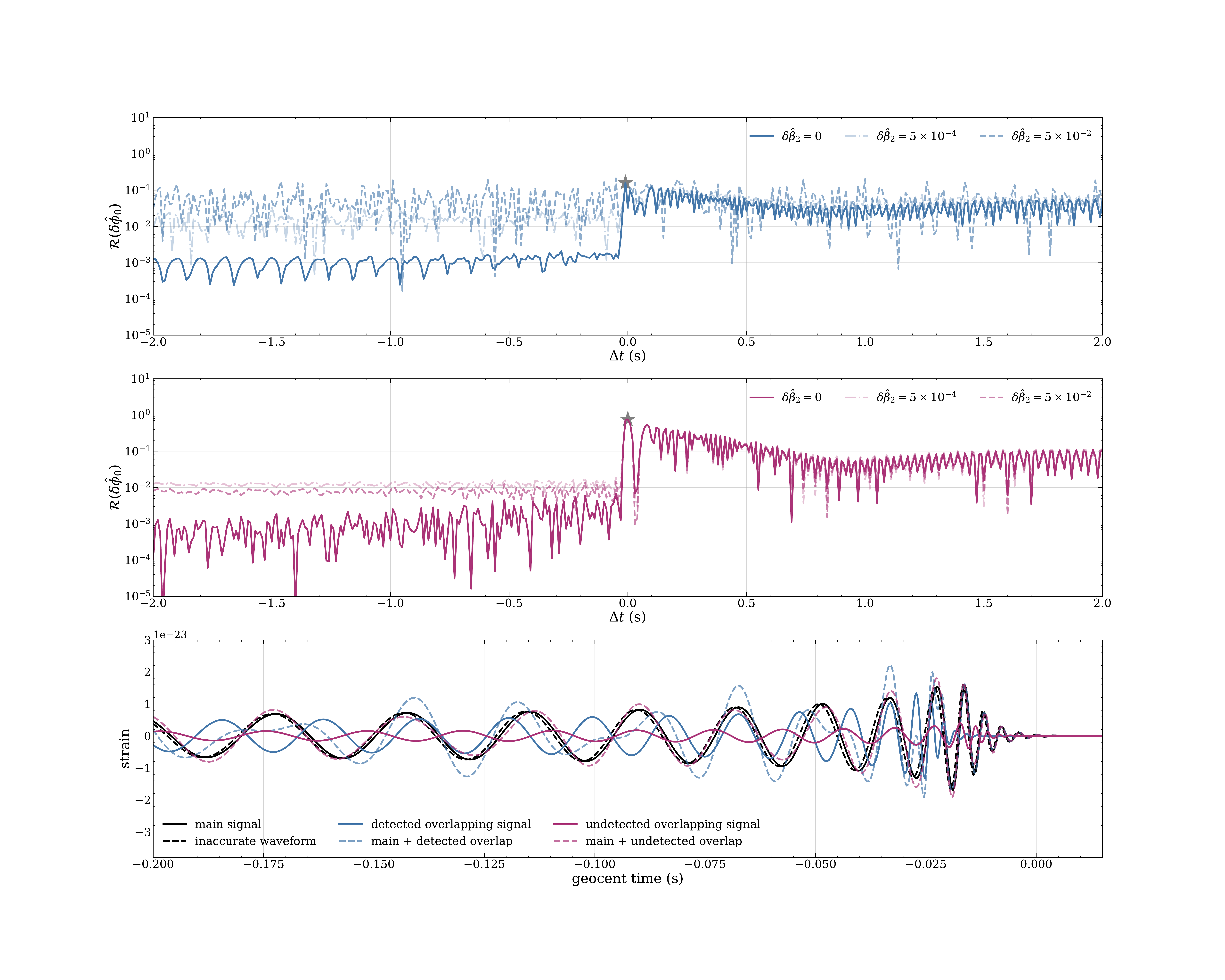}
    \centering
    \caption{\label{pic:example_event} Uppermost and middle rows: The error ratio of $\delta \hat{\phi}_0$ varies with merger time difference. The main signal has (detector frame) $\mathcal{M}_\mathrm{c} = 32\,\mathrm{M}_\odot$, $q=0.9$, $\chi_\mathrm{eff}=0.2$ and SNR of 27. The overlapping signal is an equal mass BBH with $\mathcal{M}_\mathrm{c} = 20\,\mathrm{M}_\odot$ and $\chi_\mathrm{eff}=0.1$. The SNR of the overlapping signal is adjusted by changing its luminosity distance: the detected overlap is shown in upper panel, and the undetected in the lower one. We use three kinds of waveforms explained in Sec.~\ref{sec2b}: perfect waveform (solid line), ``current'' waveform (dashed line), and ``future waveform'' (faint dotted-dashed line).\\{Bottom row: waveforms of the main and overlapping signals and their superposition. Merger times of overlapping signals are chosen to maximize their influences, as marked by grey stars in the first two rows. Inaccurate waveform in the $\delta \hat{\beta}_2=5\times10^{-2}$ case is also plotted for comparison.}} 
\end{figure*}

The error ratio for this example event is shown in Fig.~\ref{pic:example_event}, {including an illustration of the waveforms.} The error from the overlapping signal oscillates when $\Delta t$ changes due to the repeating alignments and misalignments of phases of the two GWs. The overlap error is not symmetric around $\Delta t=0$ because the two waveforms of are not symmetric, but the peak is always located in the region $|\Delta t|\leq 1$s, meaning the overlapping signal only produces a large influence when two mergers are very close. {Waveforms in the last row show how the main signal is modulated by overlapping signals.  Around $\Delta t \sim 0$, the confusion signal has larger impacts than waveform systematics, so it dominates the systematic error. The detected signal changes the signal significantly, but it is then subtracted from data and therefore produces less residual strains}. When $|\Delta t|$ is large, it is waveform inaccuracy that dominates the systematic error. 
These characteristics are consistent with previous works~\citep{himemoto2021_ImpactsOverlappingGravitationalwave,samajdar2021_BiasesParameterEstimation,relton2021_ParameterEstimationBias,antonelli2021_NoisyNeighboursInference}. 

\begin{figure}
    \includegraphics[width=0.45\textwidth]{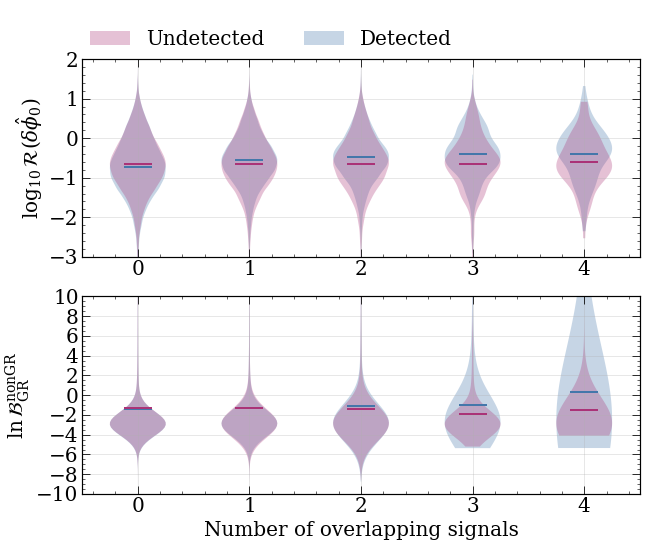}
    \centering
    \caption{\label{pic:violin} {Distributions of systematic errors and Bayes factors in the high merger rate catalog with $\delta \hat{\beta_2} = 5\times 10^{-2}$ waveform, classified by number and type of overlapping signals. Bars denote the mean value. The number of overlapping signals is cut at 4 because of the insufficient number of events coming with $>4$ overlapping signals.
    The difference in the increase of mean values shows detected overlapping signals could magnify the effects of inaccurate waveform models.} }
\end{figure}

It is possible for undetected signals to produce significant systematic errors in our simulation. However, comparing the detected and undetected overlapping signal, the former produces larger systematic error when the waveform is not accurate because the waveform systematic is also involved in signal subtraction. {One can see this from the more intense perturbations in $\delta \hat{\beta}_2 \neq 0$ case in Fig.~\ref{pic:example_event} for detected overlaps (errors are directly added so they may constructively or destructively interfere.) This implies different types of systematic errors are correlated and could be a magnifying factor for each other, as expected from Eq.~\ref{eq:do}. A more direct comparison is given in Fig.~\ref{pic:violin}. We calculate systematic errors for each BBH event in our mock catalog, and show systematic errors and Bayes factors caused by different numbers of overlapping signals for the high merger rate catalog. With the increase of the number of overlaps, detected overlaps tend to produce larger errors, while errors from confusion background signals make smaller incremental changes.}

\begin{figure*}
    \includegraphics[width=0.9\textwidth]{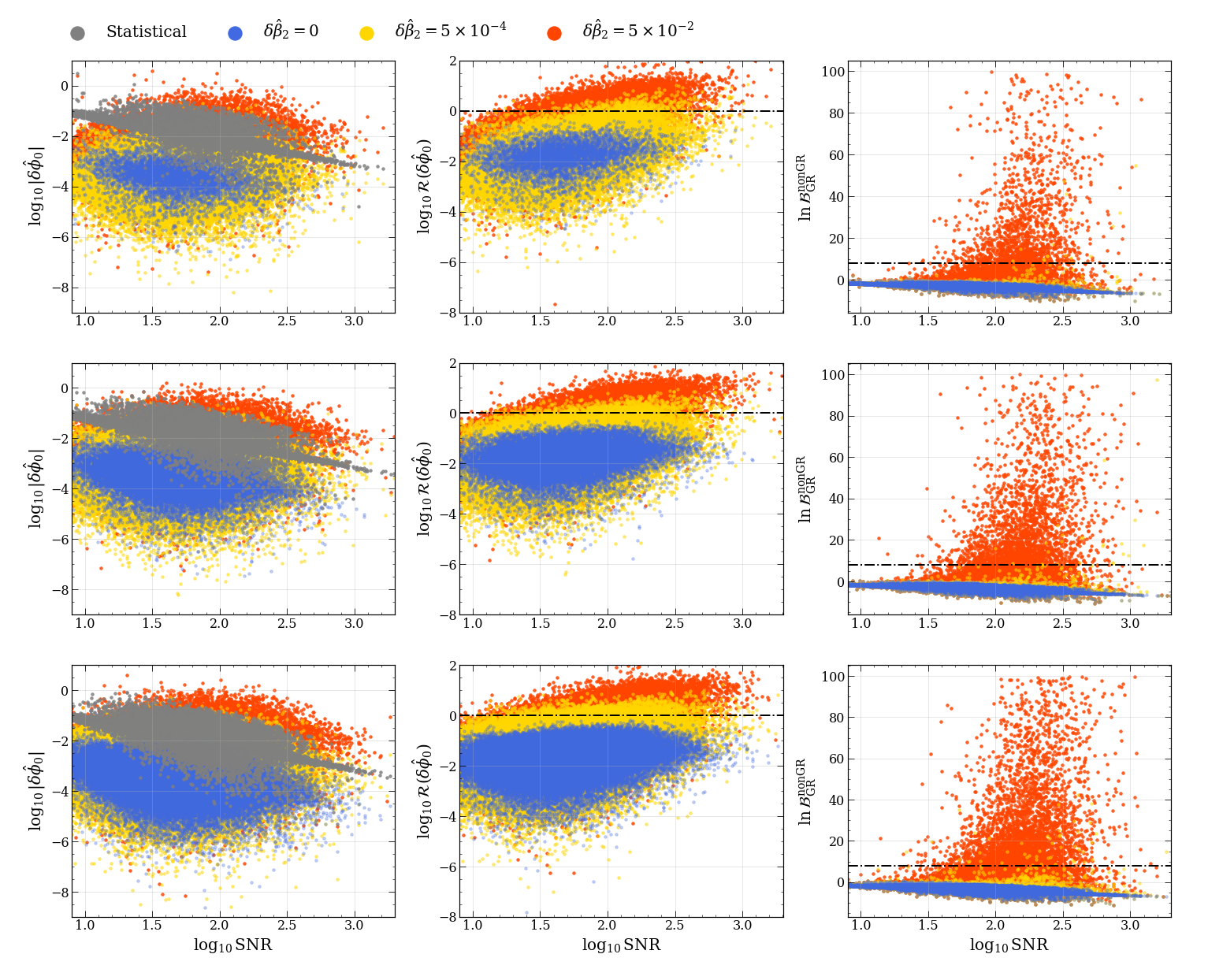}
    \centering
    \caption{\label{pic:error_vs_snr} Relation between SNR and absolute error (first column), error ratio (second column) of $\delta \hat{\phi}_0$ and Bayes factor (third column) for low (uppermost row), median (median row), and high (bottom row) merger rate catalogs. Each point represents a BBH event. Blue points are for the ``perfect waveform'' case, where all systematic errors come from overlapping signals; red points stand for ``current waveform'' case and yellow points for ``future waveform'' case. Grey points in the first column are statistical errors. Dashed lines in the second and third columns are the threshold above which GR is mistakenly disfavoured.
    This plot shows $\mathcal{R}(\delta \hat{\phi}_0)>1$ and $\ln \mathcal{B}^{\mathrm{nonGR}}_{\mathrm{GR}}>8$ are mostly from ``current waveform'' and high SNR events. } 
\end{figure*}

The statistical error $\Delta \theta^i_\mathrm{stat} = (\bm{\Gamma}^{-1})^{ij} (\partial_j h | n) \approx 1/|h| \approx 1/\mathrm{SNR}$, while the systematic error $\Delta \theta^i_\mathrm{stat} = (\bm{\Gamma}^{-1})^{ij} (\partial_j h | \delta H)$ does not necessarily shrink when the SNR increases, for example, waveform systematics of the main signal. Therefore, systematic errors may dominate in high SNR scenario. We plot the absolute error, error ratio {and Bayes factor} with SNR in Fig.~\ref{pic:error_vs_snr}. {We find that the} error ratio could exceed one for the ``current'' waveform, and this happens more often when SNR$>30$ despite the fact that high SNR events are rarer. Error ratios for the ``future'' waveform simulations are usually below one, but a certain amount of exceptions exist. {For the Bayes factor analysis we find a similar situation, although there are a smaller fraction of more extreme values. There are roughly 0.8\% and 18\% events producing $\mathcal{R}(\delta \hat{\phi}_0)>1$ for $\delta \hat{\beta}_2=5\times10^{-4}$ and $5\times10^{-2}$, respectively, while for $\ln \mathcal{B}^{\mathrm{nonGR}}_{\mathrm{GR}}>8$ the fractions are 0.02\% and 3\%.} As pointed out by \citet{moore2021_TestingGeneralRelativity}, false deviations could be achieved even though estimations for individual events are generally accurate. We will investigate this in more detail in the next subsection. 

\subsection{Error accumulation in a catalog\label{sec3b}}
\begin{figure*}
    \includegraphics[width=\textwidth]{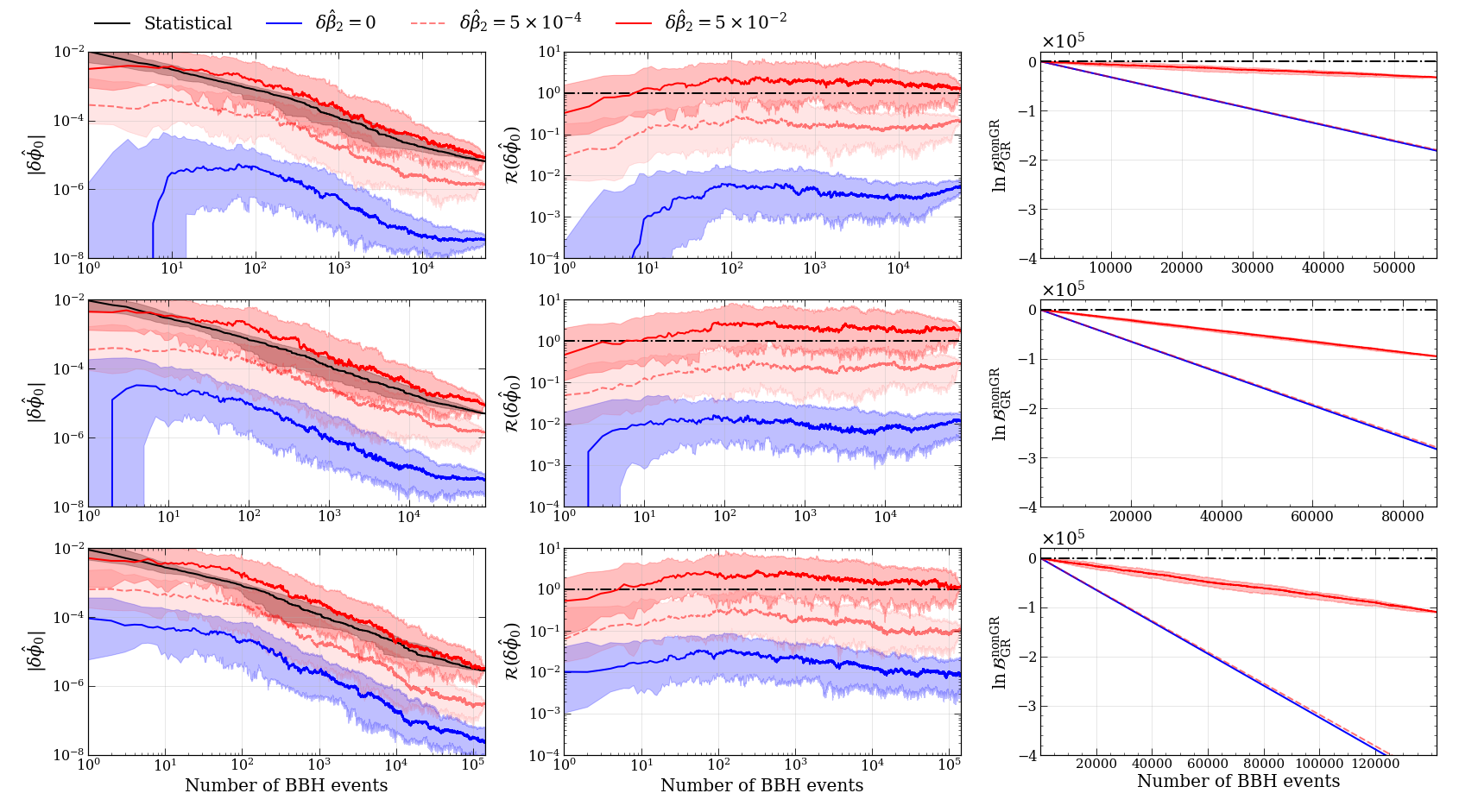}
    \centering
    \caption{\label{pic:error_accu_all} Systematic error accumulates with the increase of number of events. The first column shows the absolute error of $\delta \hat{\phi}_0$ and the second column shows the error ratio. The third column is the Bayes factor. Solid red lines are ensemble average for ``current waveform'', dashed red lines are for ``future waveform'', and blue lines stand for the perfect waveform. The shadow along lines are 68\% confidence interval. The first, second, and third rows are for low, median, and high catalogs, respectively. The black dotted-dashed line is the threshold above which a false deviation of GR is claimed.
    {False deviations can be achieved with the increase in the number of events by multiplying posteriors, but multiplying Bayes factors does not give wrong conclusions in these correct-analyses-dominated catalogs.} } 
\end{figure*}
As mentioned in Sec.~\ref{sec2d}, we combine all BBH events by multiplying likelihoods or Bayes factors. The results are shown Fig.~\ref{pic:error_accu_all}.
Let $N_\mathrm{event}$ be the number of events. When multiplying likelihoods, the statistical uncertainty shrinks as $1/\sqrt{N_\mathrm{event}}$. The absolute error of the testing parameter also decreases, but at a slower pace due to the perturbations from newly accumulating systematic errors. It also follows $1/\sqrt{N_\mathrm{event}}$ if there were no systematic errors - we observe that the test with the perfect waveform in a low merger rate catalog is approximately doing so. In most simulations it is the waveform inaccuracy that keeps contributing to the systematic errors. The slower decay of systematic error results in a climbing error ratio as the number of events increases. At some point (typically $\sim 10^3$ events, considering error bars) it leads to a false deviation of GR for the ``current'' waveform. For the better waveform, the error ratio climbs as well, but it keeps below the statistical level until $10^5-10^6$ events.

{Multiplying Bayes factors is a direct addition of $\ln \mathcal{B}^{\mathrm{nonGR}}_{\mathrm{GR}}$. ``Correct analyses'' can effectively decrease the combined Bayes factor so that a correct-analyses-dominated catalog leads to correct conclusions. Since there are only 3\% of events with $\ln \mathcal{B}^{\mathrm{nonGR}}_{\mathrm{GR}}>8$ (furthermore, only 7\% of events with $\ln \mathcal{B}^{\mathrm{nonGR}}_{\mathrm{GR}}>0$) for the current waveform, the sum of all Bayes factors is negative, thus false deviation is not achieved in this case. In contrast, multiplying likelihoods linearly adds systematic errors: for Gaussian distributions $f$ and $g$, the mean of their product is $\mu_{f g}=\frac{\mu_f \sigma_g^2+\mu_g \sigma_f^2}{\sigma_f^2+\sigma_g^2}$. Correct analysis and different sign of errors could diminish systematic error a bit, 
but it is never guaranteed for the error to be held around 0. Moreover, statistical uncertainty also shrinks during events stacking, so the error ratio shows a clear increase.}

%Multiplying Bayes factors, however, is more sensitive to systematics errors in PE because of the quadratic term. The ``current'' waveform could lead to a strong false deviation of GR with only tens of events. Intriguingly, this method differs from multiplying likelihoods in the behavior of tests with ``future'' waveform in the high merger rate catalog. Multiplying likelihoods does not lead to false deviation in this case, but multiplying Bayes factors finally claims deviation from GR after some oscillations at first $\sim 5000$ events. Compared with the lower merger rates, we conclude the frequent inaccurate subtractions of detected overlapping signals amplify the effects of waveform systematics, even if we have used a relatively accurate waveform model with mismatches around $10^{-7}-10^{-6}$.  

\subsection{Golden events\label{sec3c}}
We have combined all the detected BBH events in the above subsection. It is also interesting to test GR with only the ``golden events'', i.e., the GW events with high SNR and clean data that contribute to most of the information in the whole catalog test. This idea is widely used in many works, such as recent GWTC-3 tests of GR~\citep{LIGOScientific:2021sio} and cosmology~\citep{LIGOScientific:2021aug}. Since the noise is Gaussian in our simulation, we select the golden events with only two criteria: SNR above a chosen threshold (50 or 200) and that there is no detected overlapping signals. 

\begin{figure*}
    \includegraphics[width=1\textwidth]{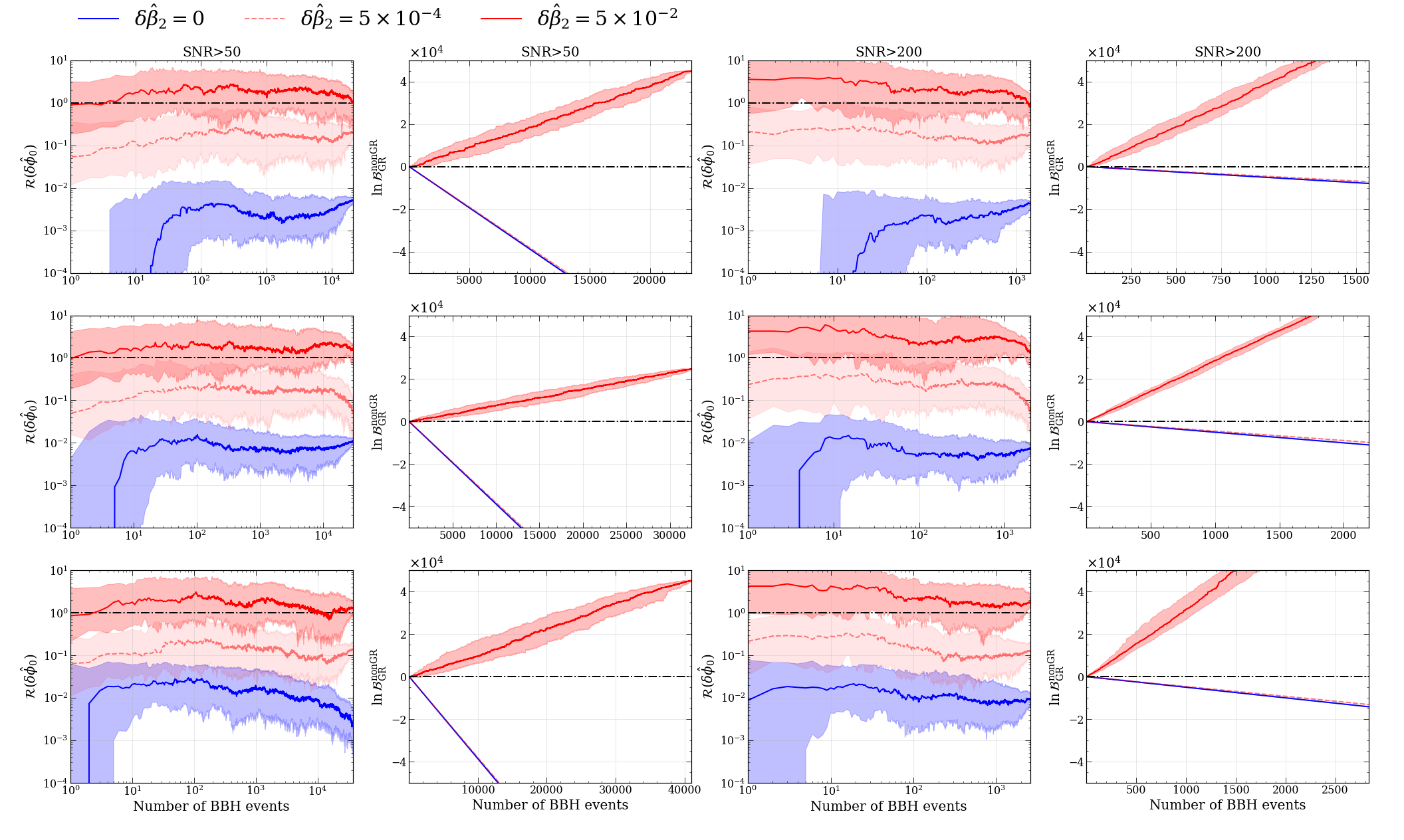}
    \centering
    \caption{\label{pic:golden_events} Similar to Fig.~\ref{pic:error_accu_all}, the error ratio and Bayes factor accumulation. The left two columns show results from SNR$>$50 events, the right two columns are for SNR$>$200 events. Compared with Fig.~\ref{pic:error_accu_all}, it shows that tests with high SNR events are more likely to make a false deviation from GR.} 
\end{figure*}

Results for the error ratio and Bayes factor are shown in Fig.~\ref{pic:golden_events}: high SNR events are more vulnerable to systematic errors. Fewer events are needed to create a false deviation for the ``current'' waveform model, and the ``future'' waveform is closer to false deviation in all three catalogs. {Moreover, the golden events catalog consists of more incorrect analyses ($\mathcal{R}(\delta \hat{\phi}_0)>1$ or $\ln \mathcal{B}^{\mathrm{nonGR}}_{\mathrm{GR}}>8$), and it causes the Bayes factor of current waveform to incorrectly favor the non-GR theory.}

As mentioned in Sec.~\ref{sec3a}, statistical uncertainty decreases as $1/\mathrm{SNR}$ while systematics do not as long as waveform is imperfect. The false deviation for golden events is not surprising from this angle, but it does need more attention and an appropriate solution for future data analysis.

\section{Conclusions and discussions\label{sec4}}
We have investigated how systematic errors in testing GR accumulate under the influence of overlapping signals and inaccurate waveforms. We have considered different levels of waveform inaccuracies and event rates, and employed two approaches to combining the results. 

We confirm that systematic errors could accumulate when combining multiple events, and could lead to incorrectly disfavoring GR in some cases. Since overlapping signals do not always occur, it is waveform inaccuracies that keep contributing to the systematic error in the catalog tests. An accurate waveform model is effective at preventing false deviations in most cases, while a worse one could lead to biased conclusions. We additionally find that overlapping signals can enlarge the effect of waveform systematics. {By increasing the number of overlaps, we tend to achieve a greater systematic error and a Bayes factor that leans more toward the non-GR model.} One can avoid this correlated error by selecting events with no detected overlapping signals, and, if one prefers, with high SNR as well. However, we have showed these events produce biases much faster because waveform systematics dominate in high SNR scenario. 

{We should point out that GR is assumed to be the true theory to describe the data in this work, which is not necessarily correct. The inverse problem, namely, what happens to detection and PE when we use GR waveform for data analysis but GR is wrong (stealth bias), is investigated in previous works~\citep{Cornish:2011ys,vallisneri2013_StealthBiasGravitationalWave,vitale2014_HowSeriousCan}.  The core idea of our work and stealth bias is the same: using an incorrect model in data analysis can lead to biased results. Stealth bias emphasizes the importance of assuming the correct theory, while our work points out that even if the assumed fundamental theory is correct, waveform modelling and overlapping signals are still able to corrupt the results.}

We re-emphasize that systematic errors can accumulate when combining multiple events and lead to incorrect scientific conclusions. This problem is universal: in addition to tests of GR, any analysis based on a GW catalog is faced with this issue, such as constraints on cosmological models, neutron star models \citep{PhysRevD.105.L061301}, and astrophysical population inference. Furthermore, there are more sources of systematic errors than those investigated in this work: instrumental calibration~\citep{Sun:2020wke, Hall:2017off}, glitches~\citep{Powell:2018csz,Pankow:2018qpo}, missing physical effects~\citep{Pang:2018hjb,Saini:2022igm} and so forth. A full analysis of these contributions, and their relative importance, will be essential in designing analysis strategies for 3G detectors. 
An obvious solution to these issues is continuing improvements to waveform model accuracy and instrumental stability, but we believe more efforts are needed from the angle of data analysis. A proper estimate of confusion background may be necessary \citep{Reali:2022aps}, and new techniques might be needed, such as accounting for waveform systematic errors during PE~\citep{moore2014_NovelMethodIncorporating}, performing specific analysis of residual strain~\citep{Dideron:2022tap}, and so forth.

\begin{acknowledgments}
The authors would like to thank Chris Messenger and Christian Chapman-Bird for helpful discussions and suggestions. We are grateful for computational resources provided by Cardiff University, and funded by STFC grant ST/I006285/1. QH is supported by CSC. JV is supported by STFC grant ST/V005634/1.
\end{acknowledgments}

\bibliography{refs.bib}
%\bibliography{refs}{}
\bibliographystyle{aasjournal}

\end{document}